\documentclass[a4paper,twoside]{article}
\newcommand{\affil}[1]{$^{\rm #1}$}
\usepackage{graphicx}

\date{} 
\title{\large\bf\flushleft An Accretion Disc Model For Algol
Type Eclipsing Binary System: AV
   Del}
\author{\parbox{\textwidth}{\flushleft
\vspace{-0.5cm}
{\it S.M.R. Ghoreyshi\affil{a}, J. Ghanbari\affil{a,b}, and F. Salehi\affil{b}}\\
\vspace{0.4cm}
{\small \affil{a}\,Department of Physics, School of Sciences,
Ferdowsi University of Mashhad, Mashhad,Iran}\\
{\small \,smrgho@gmail.com} \\
{\small \affil{b}\,Department of Physics, Khayyam Institute of Higher Education, Mashhad, Iran}\\
 {\small \,ghanbari@ferdowsi.um.ac.ir} \\
 {\small \,fsalehi@wali.um.ac.ir}}}
\begin{document}
\twocolumn[
\begin{changemargin}{.8cm}{.5cm}
\begin{minipage}{.9\textwidth}
\vspace{-1cm}
\maketitle
%
%
\small{\bf Abstract:} The study is to inspect the light and
radial-velocity curves of the eclipsing binary AV Del. In comparison
to other studies already done, the study shows that the absolute
elements, fundamental orbital and physical parameters of the system
are determined by using the Wilson-Devinney code. Using these
parameters, the configuration of the system is presented. Then, an
accretion disc model for the system is introduced by using the
SHELLSPEC code. The results indicate that AV Del is a semi-detached
system in which an optically thick accretion disc is surrounding the
primary star. The outer radius of the disc is 8.0 $R_\odot$,
corresponding to a distance of 1.1 $R_\odot$ from the surface of the
secondary. Also, the temperature of $T$=5700 K for the disc is
calculated.


\medskip
\medskip
\end{minipage}
\end{changemargin}
]
\small

\section{Introduction}
AV Del ($\alpha_{2000}$ = $20^h45^m31^s.47$, $\delta_{2000}$ =
$11^\circ10'26''.4$, \emph{V}=11.8) is a rather neglected variable
star which is discovered by Hoffmeister (1935) and is given the
original designation 184.1930 to it. It is a suspected member of the
cool Algols (Popper, 1996) and was reported by the author as a
double-lined system with a period of 3.85 days and a mean spectral
type of G5, although other sources have classified it as F8 (e.g.,
Halbedel 1984). The rare class of cool Algol systems are
distinguished from the classical Algol systems in that the
mass-gaining component is a late-type star rather than a B- or A-
type star. Mader et al. (2005) presented a new spectroscopic and
\emph{BVRI} photometric observations of AV Del. They determined the
radial-velocities for both components, using the two-dimensional
cross-correlation technique TODCOR (Zucker \& Mazeh 1994). To
present a detailed radial velocity and light curve analysis of the
optical data, Mader et al. (2005) combined all light curve data with
the spectroscopic observations and analyzed them using the
Wilson-Devinney (WD) code (Wilson \& Devinney 1971; Wilson 1979,
1990). They showed the system to be most likely semi-detached, with
the less massive and cooler star filling its Roche lobe.

When a star (of a binary star) in a circular orbit expands to fill
and then overfill its Roche lobe, it will start to lose mass to its
companion. Whether or not the mass transfer can proceed in a steady
stable manner depends largely on the relative rates of change of
stellar radius, and Roche lobe radius, with respect to changes of
the loser's mass. Also, it depends on a steady overfilling of the
Roche lobe which can be influenced, e.g., by the magnetic field
(magnetic activity cycle) of the cool component.

Matter that leaves the surface of one component of a binary star can
be partly or wholly accreted by the companion. The matter can
directly strike the gainer, or it can be stored in an accretion
around it, and a part of the matter may leave the system. The
question of which possibility is realized mainly depends on the
orbital period and the separation of the components of the system.
In fact, the disc may be caused by a binary star interaction. The
study of accretion discs is one of the most active areas in the
stellar astrophysics.

Many objects that involve accretion discs are to be accompanied by
bipolar jets apparently emerging from the central region normal to
the disc. Also, in the place where the gas stream from the secondary
strikes the edge of the disc, a bright spot may be formed. Moreover,
a thin shell of gas may surround the primary star that usually
because of the stellar wind of the primary expands.

All of these effects can play an important role in the evolution of
the binary systems. Therefore, studying and deliberating of the
existence and the structure of these effects can give us good
insights about interacting binary systems.

As mentioned earlier, the Algol-type binaries are interacting
systems consisting of a hot, usually more massive, main-sequence
star and a cool giant or subgiant that fills its Roche lobe. The
secondary is losing the mass to its companion, and we are observing
the circumstellar material in the form of a gas stream, a jet, a
shell, an accretion annuals, or an accretion disc. The most recent
paper that summarizes our current knowledge of the Algols is the
work by Richards \& Albright (1999).

As we see in section 2, there are evidences for a mass transfer in
AV Del. The paper tries to explore the existence of the mass
transfer and the nature of the system. First, it presents the
analysis of the system using the latest version (Van Hamme \& Wilson
2007) of WD code in section 3 and shows a new light and
radial-velocity curves analysis, and then determines physical and
absolute parameters of the system. The study is replicated without
following Mader et al's (2005) because it should use the latest
version of WD code and should obtain more parameters of the system
with more accuracy than Mader et al's ones (2005).

Although the WD code is a great code to study the physical
parameters of the primary and secondary stars of binary systems, it
is not a suitable one to explore objects like disc, jet, spot,
stream and shell. In spite of that, they are known as objects that
probably shape in the Algol-type binaries. Although the WD code has
an adjustable parameter $l_3$ (third light), its equations are not
complete for a detailed study of objects are listed above.
Therefore, the SHELLSPEC code (Budaj \& Richards 2004) is used which
is a strong program to analyze those stars which have accretion disc
and circumstellar material. Analysis of the system by using the
SHELLSPEC code is explained in section 4.

\section{Assumptions}

The paper considered \emph{B}, \emph{V}, \emph{R} and \emph{I}
filter observations of Mader et al. (2005). The latest version of WD
code was applied to solve the light and radial-velocity curves.
After obtaining the absolute parameters of the components, the
SHELLSPEC code was used to derive the parameters of the disc.
Throughout this paper, the subscripts 1 and 2 refer to the primary
(hotter) and the secondary (cooler) components, respectively.

In the process of using WD code, for both components of the system,
bolometric linear, logarithmic and square root laws for the limb
darkening were used and the optimal results were obtained for
bolometric logarithmic limb darkening law of Klinlesmith \& Sobieski
(1970) as follows:

\begin{equation}
I=I_\circ(1-x+x \cos\theta-y \cos\theta \ln(\cos\theta)),
\end{equation}
where the limb darkening coefficients \emph{x} and \emph{y} for both
components were fixed to their theoretical values, interpolated
using Van Hamme's (1993) formula which are depicted in Table 1.

The gravity darkening exponent from Lucy (1967) and the bolometric
albedos from Rucinski (2001) were chosen for convective envelopes
(g=0.32, $A=0.5$), which are in line with the final surface
temperatures. In order to reduce the number of the free parameters,
these parameters were kept constant during all iterations. It was
assumed that the binary system has zero orbital eccentricity ($e =
0.0$) and its rotational and orbital spins are synchronous ($F_1 =
F_2 = 1.0$). Also, the black body radiation models and the stellar
atmosphere formulation for local emission on the primary and
secondary stars were used, and the optimal results were obtained for
the black body radiation models. Initially, solutions were obtained
assuming $l_3\neq0.0$, but this assumption never resulted in
satisfactory fits to the observations.

Mader et al. (2005) calculated a linear ephemeris:

\begin{eqnarray}
Min. I (HJD) = 2,450,714.34779(39) \nonumber\\+3.8534528(35)E,
\end{eqnarray}
where $E$ is the number of cycles counted from the epoch of
reference and the uncertainties are given in parentheses in units of
the last significant digits. A period study of AV Del was carried
out by Qian (2002) on the basis of fewer times of eclipses. A short
period increase of about $3.15\pm0.19\times10^{-6}$ days yr$^{-1}$
was reported by Qian (2002), while the other quadratic fit derived
by Mader et al. (2005) gives:

\begin{eqnarray}
Min. I (HJD) = 2,450,714.34771(34)
\nonumber\\+3.8534620(42)E+0.50(16)\times10^{-8}E^2,
\end{eqnarray}
which corresponds to a period change of $0.95\pm0.30\times10^{-6}$
days yr$^{-1}$. These period changes may be related to the mass
transfer presumably taking place in the system (Mader et al. 2005).
The main purpose of the paper is to check the possibility of the
presence of an accretion disc in AV Del by applying model analysis
to the existing light curves.

\begin{table*}
\begin{center}
\caption{The limb darkening coefficients for AV Del}
\begin{tabular}{@{}crrrrrrrr@{}}
\hline
& & & & & & & & \\
Parameters & $x_1(bol)$ & $x_2(bol)$ & $y_1(bol)$ & $y_2(bol)$ & $x_1$ & $y_1$ & $x_2$ & $y_2$ \\
& & & & & & & & \\
\hline
& & & & & & & & \\
Filter B & 0.639 & 0.632 & 0.225 & 0.150 & 0.739 & 0.253 & 0.813 & -0.038 \\
& & & & & & & & \\
Filter V & 0.639 & 0.632 & 0.225 & 0.150 & 0.823 & 0.165 & 0.842 & -0.232 \\
& & & & & & & & \\
Filter R & 0.639 & 0.632 & 0.225 & 0.150 & 0.647 & 0.272 & 0.750 & 0.092 \\
& & & & & & & & \\
Filter I & 0.639 & 0.632 & 0.225 & 0.150 & 0.554 & 0.265 & 0.652 & 0.150 \\
& & & & & & & & \\
\hline
\end{tabular}
\end{center}
\end{table*}

\section{WD Solutions}

The photometric data reported by Mader et al (2005) was used for
analysis by the WD code. Values were specified for the inclination
angle (\emph{i}), the mass ratio ($q\equiv$ $M_2$/$M_1$), the
modified gravitational potentials ($\Omega$, in the WD usage), the
relative monochromatic luminosity of the primary ($L_1$) and the
mean temperature of the secondary ($T_2$). The temperature of the
primary was held fixed at the spectroscopic value of $T_1$=6000K.
The monochromatic luminosity of the secondary ($L_2$) was computed
by the program directly from the temperatures of the primary and
secondary, the luminosity of the primary, the black body radiation
laws, and the geometry of the system. Other WD code parameters were
kept fix along with solutions.

Initially, values were obtained in a mode which is appropriate for
semi-detached systems following the expectation from the
characteristics of AV Del. Specifically, mode 5 was used for the
secondary which is filling its Roche lobe. The paper also tried to
use the detached configuration to complete the work. Similar to
Mader et al's results (2005), the optimal results were determined
for the semi-detached configuration.

The light curve program (LC) of WD code was implemented to employ an
initial set of values for the next step of solutions using the
differential corrections main program (DC) of WD code. After several
runs of DC program, the final set of values were obtained. As
mentioned earlier, the assumption $l_3\neq0.0$ never resulted in
satisfactory fits to the observations however a valid solution was
obtained for $l_3=0.0$. It can be inferred that the best solution is
obtained without counting for any additional effects like an
accretion disc. The results of photometric solutions with the final
elements of the system are given in Table 2. The theoretical light
curves computed with the results (continuous line) for each filter
are shown in Fig. 1.

\begin{table*}
\begin{center}
{\tiny \caption{The photometric parameters of AV Del}
\begin{tabular}{@{}lllllllllllll}
\hline \\
\multicolumn{6}{@{}c@{}}{This Work} & \multicolumn{7}{@{}c@{}}{Mader et al. (2005)} \\
\cline{1-1}\cline{2-1}\cline{3-1}\cline{4-1}\cline{5-1}\cline{6-1}\cline{7-1}\cline{9-1}\cline{10-1}\cline{11-1}
\cline{12-1}\cline{13-1}\\
Parameters & B & V & R & I & BVRI & BVRI & & B & V & R & I & BVRI \\
& Filter & Filter & Filter & Filter & Filters & $+$ \emph{RV} & & Filter & Filter & Filter & Filter & $+$ \emph{RV} \\
\hline & & \\
$i$ & 81.389 & 80.548 & 81.333 & 81.480 & 81.165 & 80.924 & & 81.06 & 81.07 & 81.42 & 81.55 & 81.344 \\
& $\pm$ 0.121 & $\pm$ 0.102 & $\pm$ 0.097 & $\pm$ 0.147 & $\pm$ 0.090 & $\pm$ 0.061 & & $\pm$ 0.20 & $\pm$ 0.12 & $\pm$ 0.10 & $\pm$ 0.16 & $\pm$ 0.066 \\
& & & & & & \\
$q$ & 0.4968 & 0.4857 & 0.4922 & 0.4611 & 0.4929 & 0.4901 & & 0.4925 & 0.4878 & 0.4874 & 0.4865 & 0.4852 \\
& $\pm$ 0.0031 & $\pm$ 0.0025 & $\pm$ 0.0028 & $\pm$ 0.0059 & $\pm$ 0.0068 & $\pm$ 0.0028& & $\pm$ 0.0042 & $\pm$ 0.0052 & $\pm$ 0.0046 & $\pm$ 0.0057 & $\pm$ 0.0035 \\
& & & & & & \\
$T_1$ & 6000 & 6000 & 6000 & 6000 & 6000 & 6000 & & 6000 & 6000 & 6000 & 6000 & 6000 \\
& fixed & fixed & fixed & fixed & fixed & fixed & & fixed & fixed & fixed & fixed & fixed \\
& & & & & & \\
$T_2$ & 4245 & 4238 & 4263 & 4324 & 4254 & 4189 & & 4195 & 4270 & 4264 & 4289 & 4275 \\
& $\pm$ 21 & $\pm$ 16 & $\pm$ 11 & $\pm$ 15 & $\pm$ 9 & $\pm$ 12 & & $\pm$ 28 & $\pm$ 15 & $\pm$ 13 & $\pm$ 18 & $\pm$ 8 \\
& & & & & & \\
$\Omega_1$ & 5.7227 & 5.6414 & 5.8148 & 5.5624 & 5.8555 & 5.6241 & & 5.71 & 5.68 & 5.54 & 5.78 & 5.581 \\
& $\pm$ 0.1275 & $\pm$ 0.0919 & $\pm$ 0.1091 & $\pm$ 0.1405 & $\pm$ 0.0378 & $\pm$ 0.6704 & & $\pm$ 0.15 & $\pm$ 0.10 & $\pm$ 0.11 & $\pm$ 0.15 & $\pm$ 0.048 \\
& & & & & & \\
$\Omega_2$ & 2.8707 & 2.8490 & 2.8597 & 2.7998 & 2.8632 & 2.8567 & & 2.861 & 2.852 & 2.852 & 2.850 & 2.847 \\
& & & & & & & & $\pm$ 0.015 & $\pm$ 0.015 & $\pm$ 0.015 & $\pm$ 0.015 & $\pm$ 0.005 \\
& & & & & & \\
$L_1)_B$ & 9.3337 & . . . & . . . & . . . & 9.1694 & 8.8915 & & 7.787 & . . . & . . . & . . . & 7.635 \\
& $\pm$ 0.011 & & & & $\pm$ 0.021 & $\pm$ 0.009 & & $\pm$ 0.011 & & & & \\
& & & & & & \\
$L_1)_V$ & . . . & 8.5210 & . . . & . . . & 8.2909 & 7.8759 & & . . . & 6.815 & . . . & . . . & 6.883 \\
& & $\pm$ 0.007 & & & $\pm$ 0.014 & $\pm$ 0.007 & & & $\pm$ 0.008 & & & \\
& & & & & & \\
$L_1)_R$ & . . . & . . . & 7.2809 & . . . & 7.2505 & 6.7359 & & . . . & . . . & 6.168 & . . . & 6.107 \\
& & & $\pm$ 0.001 & & $\pm$ 0.021 & $\pm$ 0.009 & & & & $\pm$ 0.009 & & \\
& & & & & & \\
$L_1)_I$ & . . . & . . . & . . . & 6.6582 & 6.4409 & 5.8583 & & . . . & . . . & . . . & 5.222 & 5.450 \\
& & & & $\pm$ 0.006 & $\pm$ 0.016 & $\pm$ 0.010 & & & & & $\pm$ 0.013 & $\pm$ 0.005 \\
& & & & & & \\
$L_2)_B$ & 2.7929 & . . . & . . . & . . . & 2.9321 & 3.2037 & & . . . & . . . & . . . & . . . & . . . \\
& & & & & & \\
$L_2)_V$ & . . . & 3.7063 & . . . & . . . & 4.0253 & 4.4048 & & . . . & . . . & . . . & . . . & . . . \\
& & & & & & \\
$L_2)_R$ & . . . & . . . & 4.8852 & . . . & 4.9086 & 5.3508 & & . . . & . . . & . . . & . . . & . . . \\
& & & & & & \\
$L_2)_I$ & . . . & . . . & . . . & 5.5548 & 5.7543 & 6.2337 & & . . . & . . . & . . . & . . . & . . . \\
& & & & & & \\
log g$_1$& 3.78 & 3.77 & 3.80 & 3.77 & 3.80 & 3.91 & & . . . & . . . & . . . & . . . & 3.759 \\
& $\pm$ 0.02 & $\pm$ 0.02& $\pm$ 0.02 & $\pm$ 0.02 & $\pm$ 0.02 & $\pm$ 0.02 \\
& & & & & & \\
log g$_2$ & 3.03 & 3.03 & 3.03 & 3.03 & 3.03 &  3.03& & . . . & . . . & . . . & . . . & 3.032\\
& $\pm$ 0.01 & $\pm$ 0.01& $\pm$ 0.01 & $\pm$ 0.01 & $\pm$ 0.01 & $\pm$ 0.01 \\
& & \\
$r_1(pole)$ & 0.1911 & 0.1937 & 0.1875 & 0.1957 & 0.1862 & 0.1653 & & 0.1913 & 0.1925 & 0.1974 & 0.1886 & 0.1959 \\
& $\pm$ 0.0047 & $\pm$ 0.0042 & $\pm$ 0.0042 & $\pm$ 0.0054 & $\pm$ 0.0017 & $\pm$ 0.0183 \\
& & & & & & \\
$r_1(point)$ & 0.1933 & 0.1960 & 0.1895 & 0.1980 & 0.1881 & 0.1665 & & 0.1935 & 0.1947 & 0.1999 & 0.1906 & 0.1983 \\
& $\pm$ 0.0049 & $\pm$ 0.0044 & $\pm$ 0.0044 & $\pm$ 0.0057 & $\pm$ 0.0018 & $\pm$ 0.0188 \\
& & & & & & \\
$r_1(side)$ & 0.1921 & 0.1947 & 0.1884 & 0.1968 & 0.1871 & 0.1659 & & 0.1923 & 0.1935 & 0.1985 & 0.1895 & 0.1970 \\
& $\pm$ 0.0048 & $\pm$ 0.0043& $\pm$ 0.0043 & $\pm$ 0.0055 & $\pm$ 0.0017 & $\pm$ 0.0185 \\
& & & & & & \\
$r_1(back)$ & 0.1931 & 0.1957 & 0.1892 & 0.1977 & 0.1879 & 0.1664 & & 0.1932 & 0.1944 & 0.1996 & 0.1904 & 0.1980 \\
& $\pm$ 0.0049 & $\pm$ 0.0044 & $\pm$ 0.0044 & $\pm$ 0.0056 & $\pm$ 0.0018 & $\pm$ 0.0188 \\
& & & & & & \\
$r_2(pole)$ & 0.2993 & 0.2976 & 0.2984 & 0.2935 & 0.2987 & 0.2982 & & 0.2986 & 0.2978 & 0.2978 & 0.2976 & 0.2974 \\
& $\pm$ 0.0021 & $\pm$ 0.0020 & $\pm$ 0.0017 & $\pm$ 0.0027 & $\pm$ 0.0011 & $\pm$ 0.0004 \\
& & & & & & \\
$r_2(point)$ & 0.4287 & 0.4264 & 0.4275 & 0.4211 & 0.4279 & 0.4272 & & 0.4277 & 0.4268 & 0.4267 & 0.4265 & 0.4262 \\
& $\pm$ 0.0085 & $\pm$ 0.0083 & $\pm$ 0.0071 & $\pm$ 0.0112 & $\pm$ 0.0045 & $\pm$ 0.0018 \\
& & & & & & \\
$r_2(side)$ & 0.3124 & 0.3105 & 0.3115 & 0.3061 & 0.3118 & 0.3112 & & 0.3116 & 0.3108 & 0.3107 & 0.3106 & 0.3104 \\
& $\pm$ 0.0022 & $\pm$ 0.0021 & $\pm$ 0.0018 & $\pm$ 0.0029 & $\pm$ 0.0012 & $\pm$ 0.0005 \\
& & & & & & \\
$r_2(back)$ & 0.3449 & 0.3430 & 0.3440 & 0.3387 & 0.3443 & 0.3437 & & 0.3441 & 0.3433 & 0.3433 & 0.3431 & 0.3429 \\
& $\pm$ 0.0022 & $\pm$ 0.0021 & $\pm$ 0.0018 & $\pm$ 0.0028 & $\pm$ 0.0011 & $\pm$ 0.0005 \\
& & & & & & \\
$\Omega_{in}$ & 2.8707 & 2.8490 & 2.8597 & 2.7998 & 2.8632 & 2.8567 & & . . . & . . . & . . . & . . . & . . . \\
& & & & & & \\
$\Omega_{out}$ & 2.5735 & 2.5579 & 2.5656 & 2.5224 & 2.5681 & 2.5635 & & . . . & . . . & . . . & . . . & . . . \\
& & & & & & \\
$\sigma$ & 0.03 & 0.025 & 0.020 & 0.026 & 0.028 & 0.567 & & . . . & . . . & . . . & . . . & . . . \\
& & & & & & \\
\hline
\end{tabular}}
\end{center}
\end{table*}


\begin{figure*}
\begin{center}
\includegraphics[width=1.0\textwidth]{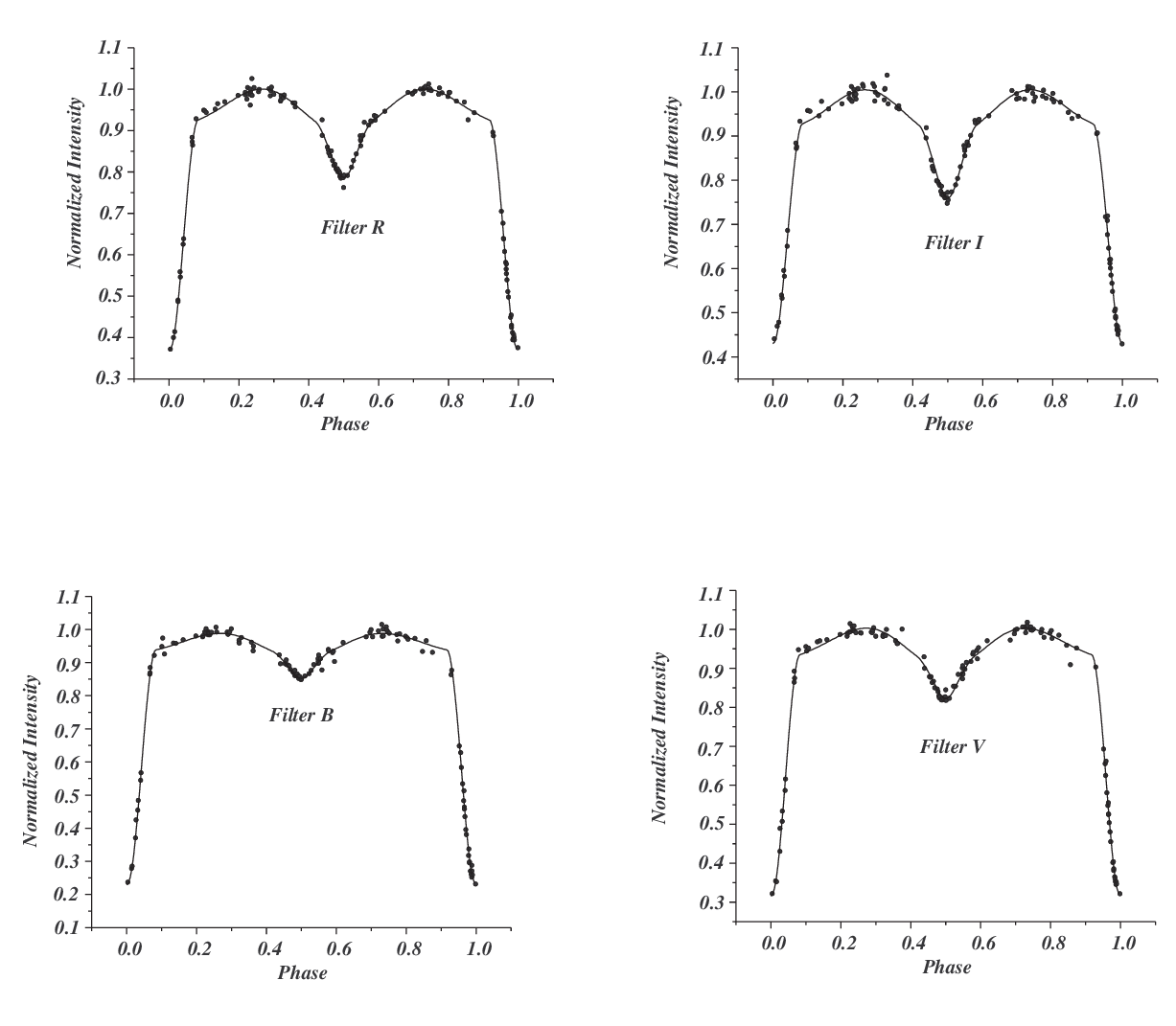}
\caption{The observed and theoretical light curves of AV Del. The
theoretical light curves are drawn on the basis of the values
derived via WD code. Solid circles indicate the observed data and
the continuous lines show the theoretical light curves. Up right and
left diagrams clarify \emph{I} and \emph{R} filters, respectively.
Also, down right and left diagrams show \emph{V} and \emph{B}
filters, respectively.}
\end{center}
\end{figure*}


Using the radial-velocity data from Mader et al. (2005), the orbital
elements of the system were obtained. Similar to photometric
solutions, the WD code was used for spectroscopic solutions. Values
are found for the orbital semi-major axis (\emph{a}) and the
radial-velocity of the binary system center of mass ($V_\gamma$).
Also, the ephemeris adopted is the same as the one specified in
equation (2). The results of spectroscopic solutions are given in
Table 3. Using the final elements of the objects, the theoretical
radial-velocity curves (continuous lines) are shown in Fig. 2. Also,
all uncertainties given in tables 2 and 3 are standard errors as
reported by the WD code.

\begin{table*}
\begin{center}
\caption{The spectroscopic parameters of AV Del}
\begin{tabular}{@{}lll}
\hline &  &  \\
Parameters & This Work & Mader \\
& & et al. (2005) \\
& & \\
\hline & & \\
$V_\gamma$ (km/s) & -64.141 & -63.14 \\
& $\pm$ 0.181 & $\pm$ 0.24 \\

$a$ (R$_\odot)$ & 13.3192 & 13.34 \\
& $\pm$ 0.0347& $\pm$ 0.09 \\
& & \\
\hline
\end{tabular}
\end{center}
\end{table*}


\begin{figure*}
\begin{center}
\includegraphics[width=.60\textwidth]{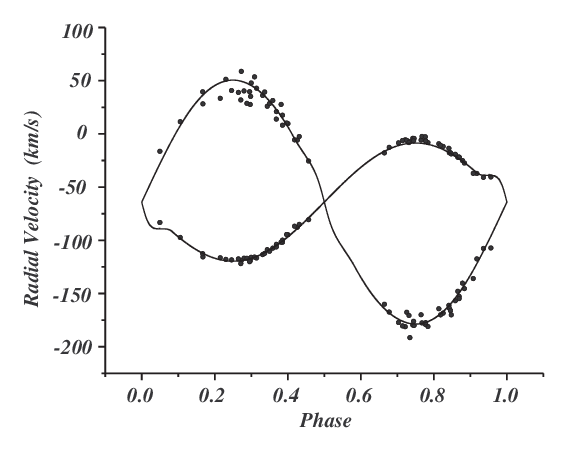}
\caption{The observed and theoretical radial-velocity curves of AV
Del. The theoretical radial-velocity curves are drawn on the basis
of the values derived by using the WD code. Solid circles show the
observed data and continuous lines indivate the theoretical
radial-velocity curves.}
\end{center}
\end{figure*}


Using the obtained results of the light and radial-velocity curves,
the absolute parameters of the system were calculated. The results
are listed in Table 4. The absolute dimensions were determined by
using the following formulae:

\begin{equation}
f_1(M_1,M_2,i)=(M_2\sin i)^3/(M_1+M_2)^2,
\end{equation}
\begin{equation}
L_{1,2}/L_\odot=(R_{1,2}/R_\odot)^2(T_{1,2}/T_\odot)^4,
\end{equation}
\begin{equation}
\rho_{1,2}/\rho_\odot=(0.01344M_{1,2})/[(M_1+M_2)P^2r^3_{1,2(side)}],
\end{equation}
$f_1$ is the mass function and $\rho_1$ and $\rho_2$ are the mean
mass density of the primary and secondary, respectively.

\smallskip
\begin{table*}
\begin{center}
\caption{The absolute elements of the binary system AV Del}
\begin{tabular}{@{}lllllllll}
\hline \\
\multicolumn{7}{@{}c@{}}{This Work} & & Mader et \\
\cline{1-1}\cline{2-1}\cline{3-1}\cline{4-1}\cline{5-1}\cline{6-1}\cline{7-1} & & & & & & & & al. (2005) \\
Parameters & B & V & R & I & BVRI & BVRI & & BVRI \\
& Filter & Filter & Filter & Filter & Filters & $+$ \emph{RV} & & $+$ \emph{RV} \\
\hline & & \\
$M_1/M_\odot$ & 1.426 & 1.437 & 1.432 & 1.462 & 1.430 & 1.433 & & 1.453 \\
& $\pm$ 0.025 & $\pm$ 0.027& $\pm$ 0.023 & $\pm$ 0.025 & $\pm$ 0.028 & $\pm$ 0.031 \\
$M_2/M_\odot$ & 0.709 & 0.699 & 0.704 & 0.674 & 0.706 &  0.702 & & 0.705 \\
& $\pm$ 0.015 & $\pm$ 0.017& $\pm$ 0.014 & $\pm$ 0.016 & $\pm$ 0.019 & $\pm$ 0.021 \\
$R_1/R_\odot$ & 2.56 & 2.59 & 2.51 & 2.62 & 2.49 & 2.21 & & 2.632 \\
& $\pm$ 0.02 & $\pm$ 0.02& $\pm$ 0.01 & $\pm$ 0.02 & $\pm$ 0.03 & $\pm$ 0.03 \\
$R_2/R_\odot$ & 4.26 & 4.24 & 4.25 & 4.18 & 4.25 & 4.24 & & 4.233 \\
& $\pm$ 0.02 & $\pm$ 0.02& $\pm$ 0.01 & $\pm$ 0.02 & $\pm$ 0.03 & $\pm$ 0.03 \\
$M_1(bol)$ & 2.58 & 2.56 & 2.63 & 2.53 & 2.64 & 2.90 & & 2.47 \\
& $\pm$ 0.04 & $\pm$ 0.04& $\pm$ 0.02 & $\pm$ 0.03 & $\pm$ 0.06 & $\pm$ 0.07 \\
$M_2(bol)$ & 2.98 & 3.00 & 2.97 & 2.94 & 2.97 & 3.04 & & 2.90 \\
& $\pm$ 0.04 & $\pm$ 0.03& $\pm$ 0.01 & $\pm$ 0.03 & $\pm$ 0.05 & $\pm$ 0.07 \\
& & \\
\hline
\end{tabular}
\end{center}
\end{table*}

The potential fillout percentage ($fillout$) for the components can
be calculated from the following formula:

\begin{equation}
fillout_{1,2}=\frac{\Omega_{in}}{\Omega_{1,2}}\times100.
\end{equation}
$\Omega_{in}$ is the critical inner potential and $\Omega_1$ and
$\Omega_2$ are the modified gravitational potentials of the primary
and secondary components, respectively.

Also, the following formulae, $K_{1,2}$ and $a_{1,2}\sin i$ were
calculated:

\begin{equation}
K_{1,2}=\frac{2\pi a_{1,2}\sin i}{P},
\end{equation}
\begin{equation}
a_1\sin i=[\frac{GP^2f_1(M_1,M_2,i)}{4\pi^2}]^\frac{1}{3},
\end{equation}
\begin{equation}
a_2\sin i=\frac{a_1\sin i}{q}.
\end{equation}
The results are tabulated in Table 5. On the basis of these values,
the configuration of the components was drawn by using the Binary
Maker 2.0 (Bradstreet, 1993) software shown in Fig. 3.

\begin{table*}
\begin{center}
\caption{Other elements of the binary system AV Del}
\begin{tabular}{@{}lllllllll}
\hline \\
\multicolumn{7}{@{}c@{}}{This Work} & & Mader et \\
\cline{1-1}\cline{2-1}\cline{3-1}\cline{4-1}\cline{5-1}\cline{6-1}\cline{7-1} & & & & & & & & al. (2005) \\
Parameters & B & V & R & I & BVRI & BVRI & & BVRI \\
& Filter & Filter & Filter & Filter & Filters & $+$ \emph{RV} & & $+$ \emph{RV} \\
& & \\
\hline
& & \\
$L_1/L_\odot$ & 7.505 & 7.682 & 7.215 & 7.861 & 7.101 & 5.593 & & . . . \\
& $\pm$ 0.092 & $\pm$ 0.081& $\pm$ 0.065 & $\pm$ 0.072 & $\pm$ 0.103 & $\pm$ 0.086 \\
$L_2/L_\odot$ & 5.207 & 5.129 & 5.257 & 5.412 & 5.227 & 4.892 & & . . . \\
& $\pm$ 0.086 & $\pm$ 0.067& $\pm$ 0.051 & $\pm$ 0.059 & $\pm$ 0.094 & $\pm$ 0.071 \\
$\rho_1/\rho_\odot$ & 0.085 & 0.082 & 0.091 & 0.081 & 0.091 & 0.133 & & . . . \\
& $\pm$ 0.001 & $\pm$ 0.001& $\pm$ 0.001 & $\pm$ 0.001 & $\pm$ 0.001 & $\pm$ 0.001 \\
$\rho_2/\rho_\odot$ & 0.010 & 0.010 & 0.010 & 0.010 & 0.010 & 0.010 & & . . . \\
& $\pm$ 0.001 & $\pm$ 0.001& $\pm$ 0.001 & $\pm$ 0.001 & $\pm$ 0.001 & $\pm$ 0.001 \\
$a_1\sin i (\times 10^6$ km/s) & 2.863 & 2.815 & 2.841 & 2.722 & 2.848 & 2.831 & & 3.015 \\
& $\pm$ 0.023 & $\pm$ 0.016& $\pm$ 0.011 & $\pm$ 0.014 & $\pm$ 0.025 & $\pm$ 0.022 \\
$a_2\sin i (\times 10^6$ km/s) & 5.757 & 5.791 & 5.780 & 5.905 & 5.773 & 5.777 & & 6.074 \\
& $\pm$ 0.023 & $\pm$ 0.018& $\pm$ 0.013 & $\pm$ 0.015 & $\pm$ 0.026 & $\pm$ 0.023 \\
$K_1$ (km/s) & 54.030 & 53.126 & 53.624 & 51.365 & 53.752 & 53.428 & & 56.90 \\
& $\pm$ 0.095 & $\pm$ 0.092& $\pm$ 0.082 & $\pm$ 0.087 & $\pm$ 0.105 & $\pm$ 0.112 \\
$K_2$ (km/s) & 108.647 & 109.290 & 1109.080 & 111.445 & 108.943 & 109.015 & & 114.64 \\
& $\pm$ 0.115 & $\pm$ 0.112& $\pm$ 0.097 & $\pm$ 0.104 & $\pm$ 0.124 & $\pm$ 0.132 \\
$f_1(M_1,M_2,i)/M_\odot$ & 0.076 & 0.072 & 0.074 & 0.065 & 0.074 & 0.073 & & . . . \\
& $\pm$ 0.001 & $\pm$ 0.001& $\pm$ 0.001 & $\pm$ 0.001 & $\pm$ 0.001 & $\pm$ 0.001 \\
$fillout$ $1$(\%) & 50.171 & 50.510 & 49.153 & 50.332 & 48.888 & 43.727 & & . . . \\
& $\pm$ 0.003 & $\pm$ 0.007& $\pm$ 0.002 & $\pm$ 0.006 & $\pm$ 0.012 & $\pm$ 0.018 \\
$fillout$ $2$(\%) & 100.000 & 100.000 & 100.000 & 100.000 & 100.000 & 100.000 & & . . . \\
& & \\
\hline
\end{tabular}
\end{center}
\end{table*}


\begin{figure*}
\begin{center}
\includegraphics[width=0.80\textwidth]{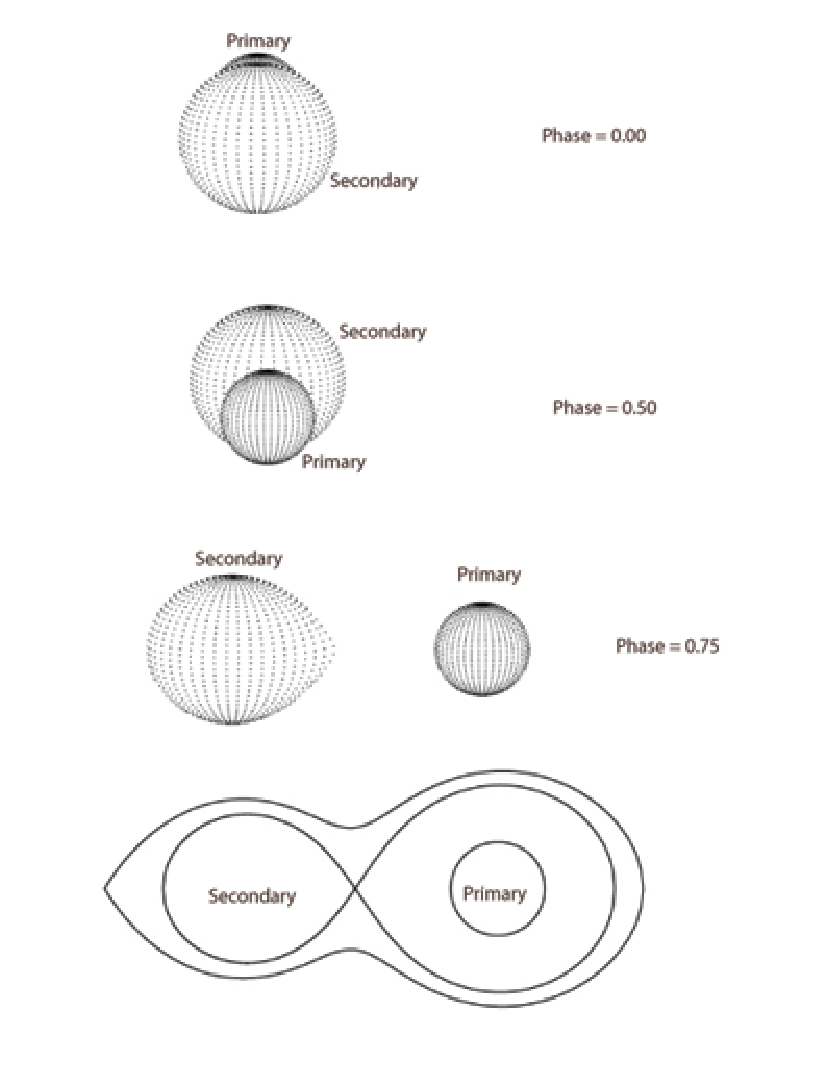}
\caption{The Configuration of the binary system AV Del on the basis
of the values derived by using the WD code. It is drawn by using the
Binary Maker 2.0 (Bradstreet, 1993) software. From up to down,
first, second and third figures show the configuration of the system
in Phases 0.00, 0.50 and 0.75, respectively. The last figure shows
the configuration of the system with the Roche lobe.}
\end{center}
\end{figure*}


\section{SHELLSPEC Solutions}

As already mentioned, the WD code is not able to study in detail the
environmental body of the binary systems, thus the assumption in
accordance with $l_3\neq0.0$ was not resulted in a good fit to the
observed data. Therefore, the SHELLSPEC code was used to solve the
light curve of AV Del to ensure whether or not the system has an
environmental body like disc, jet, stream, spot and shell. It is
believed that the study is necessary because it is predictable that
the Algol-type binaries may contain the objects. Specifically, the
secondary star of AV Del fills its Roche lobe, thus the mass
transfer probably occurs in the system and forms one or more
object(s). Besides, the distance between two components is so
sufficient that one can envisage the disc or stream to be shaped.

The SHELLSPEC code is designed to solve simple radiative transfer
along the line of sight in three-dimensional moving media. The
scattered light from a central object is taken into account assuming
an optically thin environment (Budaj \& Richards 2004). Output
intensities are then integrated through the two-dimensional
projection surface of a three-dimensional object. The assumptions of
the code include LTE and optical known state quantities and velocity
fields in three dimensions (Budaj \& Richards 2004). Using the
results obtained in section 3, the parameters of a third body were
sought in the system by using the SHELLSPEC code. In the code, both
objects the primary and the companion (secondary), may be shaped
according to Roche model used for detached, semi-detached or contact
systems. According to previous sections, in the present study, the
semi-detached configuration of the Roche model was assumed that the
primary is a nontransparent uniformly rotating sphere and the
secondary component filling its Roche lobe. The disc surrounds the
primary star and eclipses at phase 0.0. In the model, the eclipse of
the secondary star by the disc occurs at phase 0.5. The disc
thickness was assumed to be constant. Also, the black body radiation
models were employed to solutions by using the SHELLSPEC code.

The limb and the gravity darkening coefficients used in this code
are the same as the ones explained in section 2, but the reflection
effect is not included in calculations.

Since there were many free parameters in the SHELLSPEC code (because
there were several adjustable environmental body), in this stage,
the parameters of the primary and the secondary (those obtained in
section 3) were fixed. The effect of the parameters concerning to
the accretion disc, spot, jet, stream and shell on the light curve
was studied to obtain the best fit on the observed data.

Fixing the free parameters of the primary and secondary stars for
the values obtained on the basis of the WD code solutions will not
lead to the bad results because the WD code is a perfect code to
derive the parameters of the stars which are in the binary systems
and thus we do not centralize our calculations on the free
parameters of the primary and secondary and attend to the other
objects. On the other hand, \emph{R} filter observations of Mader et
al. (2005) were considered for the solutions since as it is shown in
Table 2, the accuracy of the results for this filter is more than
others.

Solutions were determined for \emph{tempsp} (effective temperature
of the spot), \emph{adisc} (half of the thickness of the disc
cylinder), \emph{rindc} (inner radius of the disc), \emph{routdc}
(outer radius of the disc), \emph{tempdc} (effective temperature of
the disc), \emph{densdc} (mean density of the disc), \emph{anedc}
(the electron number density of the disc), \emph{edendc} (the
exponent of the power-law behavior of the densities $\rho\sim$
$r^\eta$ in the disc ) and $v_{trb}$ (microturbulence). The
parameters were allowed to vary one by one and the others were kept
fixed.

First, the code was run assuming there was no third body but it
failed to obtain a good fit to the observed data. Then a disc only
was included as a third body. On the basis of the configuration,
agreement between the observed and the theoretical light curve was
very good, although the paper tried to examine the effect of other
bodies (spot, jet, stream and shell) on the light curve. Finally,
the best fit obtained while only an accretion disc was considered.
As we see in Fig. 4, when the disc is included, the agreement
between the observed data and the theoretical light curve is quite
good. It should be mentioned that the solution derived in this way
is not a unique one. of course, it shows that the fit is remarkably
improved by assuming a disc of some reliable dimensions and physical
parameters. The parameters of the disc are given in Table 6. The
configuration of the system while its primary star is surrounded by
the accretion disc on the basis of the values of Table 6 are shown
in Fig. 5.

\begin{table}[t]
\begin{center}
\caption{The parameters of the accretion disc of AV Del}
\begin{tabular}{@{}ll}
\hline & \\
Parameters & This work \\
& \\
\hline & \\
$adisc$ $(R_\odot)$ & 1.2 \\

$rindc$ $(R_\odot)$ & 3.5 \\

$routdc$ $(R_\odot)$ & 8.0 \\

$tempdc$ $(K)$& 5700 \\

$densdc$ (gr/cm$^3)$ & $33\times10^{-15}$ \\

$anedc$ (cm$^{-3})$ & $21\times10^9$ \\

$edendc$ & -1 \\

$vtrbdc$ (km/s) & 90 \\

\hline
\end{tabular}
\end{center}
\end{table}


\begin{figure*}
\begin{center}
\includegraphics[width=0.62\textwidth]{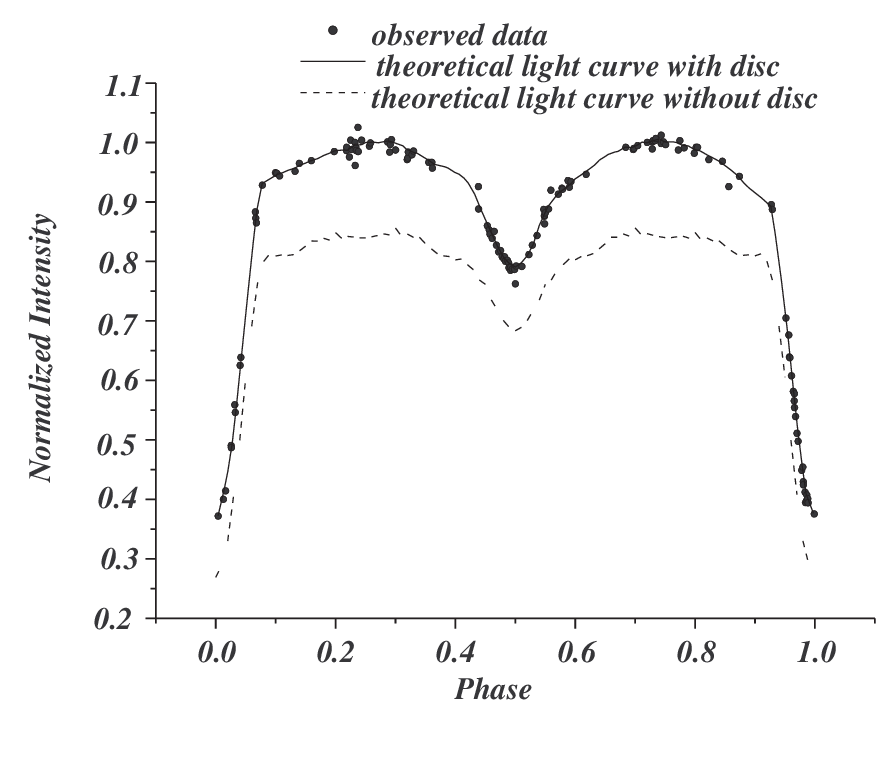}
\caption{The observed and the theoretical light curves of AV Del.
The theoretical light curves are drawn on the basis of the values
derived via SHELLSPEC code. The observed data are shown by solid
circles and the theoretical light curves are shown by continuous
(with disc) and dashed (without disc) lines.}
\end{center}
\end{figure*}


\section{Conclusion}

The latest version of WD code was applied to solve the light and
radial-velocity curves of the eclipsing binary system AV Del. The
main purpose was to check the possibility of the presence of an
accretion disc in the system. But, this tool was failed to find any
third light in AV Del system. However, the WD code was used to
determine the absolute parameters of the system. The obtained values
are very close to the ones were determined by Mader et al (2005).
The noticeable differences between the values and theirs are seen in
the combined light and radial-velocity curves (\emph{BVRI} +
\emph{RV}) solutions for $i$ $(0.42^\circ)$ and $T_2$ (86 K).
Altogether, there is no main difference between them but the present
values are more accurate than the former.

The tool called SHELLSPEC code was employed to solve the light curve
of AV Del known as an eclipsing binary system. A new configuration
for the system was presented in terms of an accretion disc
surrounding the primary star. As it is given in Table 6, the
accretion disc has the shape of a cylinder. The vertical
half-thickness (1.2 $R_\odot$) is comparable to the radius of the
primary star. The distance between the primary and the secondary
centers is about 13.3 $R_\odot$ and the outer radius of the disc is
8 $R_\odot$. Since the radiuses of the primary and the secondary are
about 2.6 $R_\odot$ and 4.2 $R_\odot$, the outer limb of the disc
will be at a distance of about 1.1 $R_\odot$ far from the surface of
the secondary. Therefore, the results are realistic. According to
the disc dimensions, it is a thick disc which the temperature is
more than that of the secondary and slightly less than that of the
primary.

Generally, on the basis of the results, AV Del had an accretion disc
while Mader et al. (2005) rejected existence of activity in the
system since they didn't find obvious signs of activity displayed by
the system. However, since there are many free parameters for the
disc, it cannot be claimed that the fit is a unique solution (Budaj
et al. 2005), but the existence of the disc in the system is
certain. Therefore, the period changes were seen in the system
(Qian, 2002) can be related to the mass transfer between two
components which has created the accretion disc.

Finally, Popper (1996) has noted although the two components appear
to have comparable luminosities, the hotter and smaller component
has the larger mass, with a mass ratio very provisionally about 1.9.
According to the values, the results are in line with his study.

According to the results, AV Del is a semi-detached system in which
the primary and secondary components have filled almost 49$\pm$1 and
100 percent of their respective critical Roche lobes.

The absolute elements of both components determine the evolutionary
state of the system. On the basis of the results, the components of
AV Del in the H-R diagram were compared with the evolutionary track
computed for the exact mass measured for the stars (Schaller et al.
1992) and for a metal abundance of Z=0.020 (Fig. 6). It appears that
AV Del includes a main-sequence primary star whereas secondary is on
the way to the giant stage.

\begin{figure*}
\begin{center}
\includegraphics[width=0.80\textwidth]{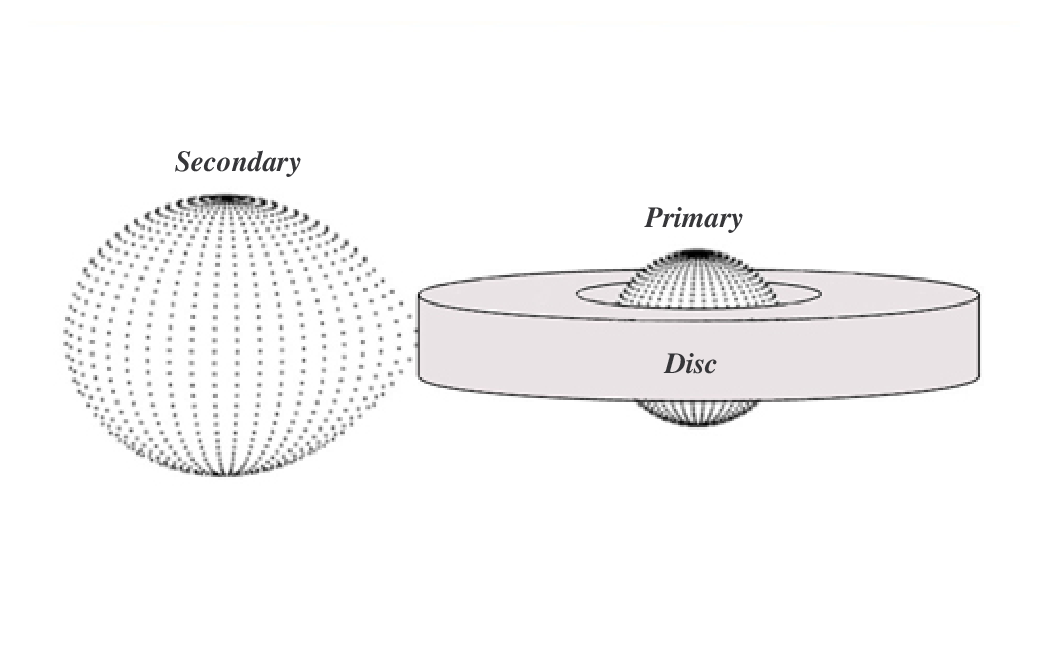}
\caption{The Configuration of the binary system AV Del on the basis
of the values derived by using the SHELLSPEC code. As we see, the
primary star is surrounded by an accretion disc.}
\end{center}
\end{figure*}


\begin{figure*}
\begin{center}
\includegraphics[width=0.80\textwidth]{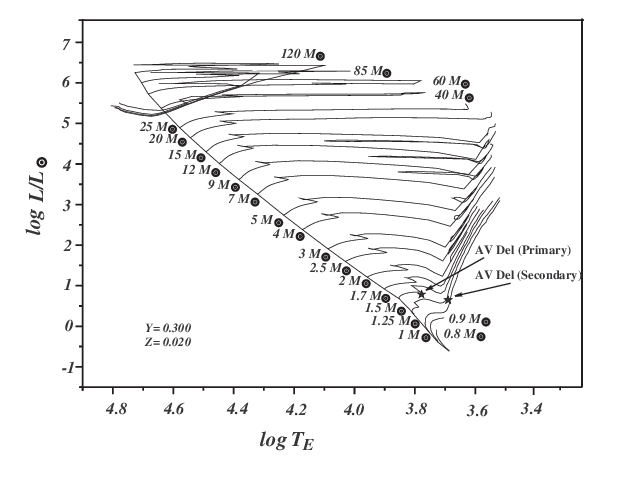}
\caption{H-R diagram for the eclipsing binary system AV Del. The
location of components are indicated with star sign. The
evolutionary track computed for the exact mass measured for the
stars (Schaller et al. 1992) and for a metal abundance of Z=0.020.}
\end{center}
\end{figure*}


\section*{Acknowledgments} We wish to thank Dr. Jano Budaj for his very useful
comments at different stages of the work and Professor Walter Van
Hamme for his very useful help for calculating the error bars of the
absolute dimensions. Also we would like to thank the referees for
their comments and suggestions.

\end{document}